\documentclass[useAMS,usenatbib]{mn2e}
\usepackage{graphicx}
\usepackage{float}
\usepackage{xcolor}
\newcommand{\aj}{AJ}			
\newcommand{\araa}{ARA\&A}		
\newcommand{\apj}{ApJ}			
\newcommand{\apjs}{ApJS}		
\newcommand{\aap}{A\&A}			
\newcommand{\mnras}{MNRAS}		
\newcommand{\pasp}{PASP}		
\newcommand{\sovast}{Soviet~Ast.}	
\newcommand{\nat}{Nature}		

\newcommand{\Ha}{H$\alpha$}
\newcommand{\Hb}{H$\beta$}

\newcommand{\Nii}{[N~{\sc ii}] $\lambda$6584}
\newcommand{\nii}{[N~{\sc ii}]}

\newcommand{\Oiii}{[O~{\sc iii}] $\lambda$5007}

\title[Dissecting galaxy triplets in the Sloan Digital Sky Survey Data Release 10]{Dissecting galaxy triplets in the Sloan Digital Sky Survey Data Release 10: I. Stellar populations and emission line analysis}
\author[Costa-Duarte M.V., O'Mill A. L., Duplancic F., Sodr\'e Jr. L., Lambas D. G.]{Costa-Duarte M. V.$^{1}$\thanks{e-mail: mvcduarte@iag.usp.br}, O'Mill A. L.$^{2,3}$, Duplancic F.$^{3,4}$, Sodr\'e Jr. L.$^{1}$, Lambas D. G.$^{2,3}$\\
$^1$ Departamento de Astronomia, Instituto de Astronomia, Geof\'\i sica e Ci\^encias Atmosf\'ericas, Universidade de S\~ao Paulo, \\ Rua do Mat\~ao 1226, Cidade Universit\'aria, 05508-090, S\~ao Paulo, Brasil.\\
$^2$ Instituto de  Astronom\'\i a Te\'orica y Experimental, IATE, Observatorio Astron\'omico, Universidad Nacional de C\'ordoba, 
\\ Laprida 854, X5000BGR, C\'ordoba Argentina\\
$^3$ Consejo Nacional de Investigaciones Cient\'ificas y T\'ecnicas (CONICET), Argentina;  Avenida Rivadavia 1917, C1033AAJ, \\ Buenos Aires, Argentina\\
$^4$ Departamento de Geof\'isica y Astronom\'ia - Facultad de Ciencias Exactas, F\'isicas y Naturales - Universidad Nacional de \\ San Juan, San Juan, Argentina}
\begin{document}

\date{Accepted . Received ; in original form }

\pagerange{\pageref{firstpage}--\pageref{lastpage}} \pubyear{2014}

\maketitle

\label{firstpage}

\begin{abstract}

We identify isolated galaxy triplets in a volume-limited sample from the Sloan Digital Sky Survey Data Release 10. Our final sample has 80 galaxy systems in the redshift range 0.04$\le$z$\le$0.1, brighter than $M_r = -20.5 + 5\log h_{70}$. Spectral synthesis results and WHAN and BPT diagnostic diagrams were employed to classify the galaxies in these systems as star-forming, active nuclei, or passive/retired.

Our results suggest that the brightest galaxies drive the triplet evolution, as evidenced by the strong correlations between properties as mass assembly and mean stellar population age with triplet properties. Galaxies with intermediate luminosity or the faintest one within the triplet seem to play a secondary role. Moreover, the relation between age and stellar mass of galaxies is similar for these galaxies but different for the brightest galaxy in the system. Most of the triplet galaxies are passive or retired, according to the WHAN classification. Low mass triplets present different fractions of WHAN classes when compared to higher mass triplets. A census of WHAN class combinations shows the dominance of star-forming galaxies in low mass triplets while retired and passive galaxies prevail in high-mass systems. We argue that these results suggest that the local environment, through galaxy interactions driven by the brightest galaxy, is playing a major role in triplet evolution. 

\end{abstract}


\section{Introduction}

%
The current inflationary cold dark matter ($\Lambda$CDM) framework predicts that galaxies and large scale structures had their origin in primordial density fluctuations in the early Universe \citep[e.g.][]{Peebles1980,WriteFrenk1991}. This hierarchical scenario of structure formation also implies that galaxies start their formation first and, as times goes by, larger and larger structures are being built by gravitational processes, such as filaments, walls, and clusters of galaxies, the so-called ``cosmic web" \citep{Zeldovichetal1982,GottIIIetal2005,AragonCalvoetal2010}. In this bottom-up scenario, groups, clusters, and superclusters of galaxies were formed in a relatively recent time when compared to galaxies themselves. Large scale structures have a strong influence on the evolution of galaxies that they harbour, i.e., galaxy properties in dense environments are statistically distinct from those in voids \citep{GroginGeller1999,Rojasetal2004}. Consequently, large scale structures can be considered as ``cosmic laboratories", being suitable objects to investigate how galaxies evolve \citep{Baloghetal1999}.

Many studies focus on galaxy clusters \citep{delaTorreetal2011} or compact groups \citep{Coendaetal2015}, due to the strong influence they have on the evolution of their galaxies. But most of the galaxies in the Universe are in smaller or less conspicuous structures, such as pairs and triplets of galaxies \citep{Karachentsev1972,Lambasetal2003}. These galaxy systems are suitable sites to investigate galaxy evolution since they are generally in low-density environments and the \textit{in situ} interaction between the members is one of the main processes in action. Galaxy mergers are an important mechanism affecting galaxies, and may lead to complex stellar populations, due to multiple starburst episodes. Indeed, a star formation enhancement is commonly reported at the end of the galaxy merger process \citep{Owersetal2007,Robothametal2013}. However, the way these galaxies evolve is not fully understood. It is also necessary to invoke internal mechanisms to reproduce some observed galaxy properties; one of them is active galactic nuclei (hereafter AGNs) feedback, which can suppress (or trigger) star formation due to a large amount of energy released to the interstellar medium \citep{Fabian2012}.

%
Recently, \cite{OMilletal2012} have compiled a catalogue of triplets of  galaxies from the Sloan Digital Sky Survey Data Release 7 \cite[SDSS-DR7,][]{Abazajianetal2007}, providing a spectroscopic and photometric sample of triplet system candidates. Further analysis carried out by \cite{Duplancicetal2013}, based on photometric colours and star formation activity, indicates that galaxy triplets present similar properties to compact groups and that these systems, comprising luminous galaxies, are actually an extension of compact galaxy groups with a smaller number of members. A dynamical analysis, employing the AA-diagram \citep{AgekyanAnosova1968} and mock catalogues, also indicates that their configurations provide better conditions for interaction and mergers between galaxy members \citep{Duplancicetal2015}.

%
A variety of techniques has been developed in order to obtain galaxy properties. Empirical spectral synthesis, by fitting an observed galaxy spectrum with a linear combination of simple stellar populations (SSPs) plus extinction attenuation, represents a powerful technique to extract integrated stellar population parameters, such as mean ages and metallicities, from a sample of galaxies \citep{Bica1988,CidFernandesetal2001,Tojeiroetal2007}.  

%
Diagnostic diagrams represent another useful tool, commonly used to investigate the mechanisms responsible for the emission lines observed in galaxy spectra. The BPT diagram \citep{BPT81} is the most used diagram for classifying emission line galaxies accordingly with the dominant photoionization source, making use of the \Oiii/\Hb\ versus \Nii/\Ha\ plane. However, emission line galaxies without significant \Oiii\ emission, such as retired and passive galaxies, cannot be distinguished from AGNs with the BPT diagram \citep[e.g., ][]{Stasinskaetal2015}. Whereas passive galaxies have negligible emission lines, retired galaxies are mainly characterized by prevailing old stellar populations and the presence of weak emission lines, which are usually explained by Hot Low-Mass Evolved Stars \citep[HOLMES,][]{Binetteetal1994}. To distinguish these low ionization galaxies from high ionization objects (AGNs), \cite{CidFernandesetal2011} have shown that an efficient way is by using the so-called WHAN diagram, i.e, the equivalent width of H$\alpha$ versus \Nii/H$\alpha$ plane. This diagram was firstly named WHAN by \cite{CidFernandesetal2011}, who introduced this acronym WHAN for simplicity (see their Introduction section).

%
The main purpose of this paper is to implement these techniques to investigate the properties of isolated triplet galaxy systems and their members. The spectral synthesis provides stellar population features of the galaxies while the diagnostic diagrams allow the classification of their main photoionization source. We will show that the information extracted from galaxy spectra with these two techniques are useful to evaluate what kind of galaxies form triplets and how their properties are correlated to the properties of their host systems. 
  
This paper is structured as follows. In Section \ref{data} we describe our volume-limited sample, presenting the spectroscopic and photometric data and the corrections applied to the data. Section \ref{triplet_identification}  presents a brief description of the selection criteria used to identify the isolated galaxy triplets in our sample. In Section \ref{spectral_analysis}, the spectral synthesis and emission line measurements are described. The general properties of galaxy triplets are presented and discussed in  Section \ref{general_configuration}. In this section, we present a concept valuable for our discussion: the luminosity hierarchy of the triplet members. Indeed, in Section \ref{mass_assembly} we present an analysis of the mass assembly, stellar mass, and age of galaxies classified according to their hierarchy and triplet stellar mass. Section \ref{elines} presents the emission line analysis by using the WHAN and BPT diagrams. Finally, our conclusions and final remarks are established in Section \ref{summary_conclusions}. Appendix \ref{appendix_spec_completeness} discuss how spectroscopic incompleteness affects the identification of galaxy triplets.

Throughout this paper, the following cosmology is adopted whenever necessary: $H_0=70 \rm\ km\ s^{-1} Mpc^{-1}$ and $(\Omega_m,\Omega_{\Lambda},\Omega_k)=(0.3,0.7,0.0)$.

\section{Data}
\label{data}

The galaxy sample analyzed in this work was drawn from the Data Release 10 of Sloan Digital Sky Survey\footnote{https://www.sdss3.org/dr10/} \citep[SDSS-DR10,][]{Ahnetal2014}. This survey covers an additional 3100 sq.deg. of the sky over the previous release and includes spectra obtained with the new spectrographs developed for APOGEE and BOSS projects. The SDSS-DR10 provides a public database of roughly two million galaxies with 5 broadband photometry ($ugriz$) and optical spectroscopy from 3800\AA\ to 9200\AA\ at the observed-frame, with a spectral resolution of R$\sim$1500 at 3800\AA. The SDSS optical spectra is a treasure trove for studies on stellar populations and on emission line properties of the galaxy populations.

The volume-limited sample employed for the detection of triplet systems of galaxy candidates at low redshift comprises galaxies brighter than $r$-band 17.77 and $M_r<-20.5+5\log h_{70}$ with spectroscopic redshift in the range 0.04$\le$z$\le$0.1. Under these constraints, our galaxy sample consists of 162,172 objects.

Notice that in this work we adopt the \textit{ModelMag} magnitude. This magnitude is estimated by fitting two distinct flux profiles and by selecting the profile with higher likelihood in the $r$-band. This magnitude is more appropriate for extended objects and also provides more robust galaxy colours. The magnitudes were corrected by extinction,  offsets \citep[values adopted from][]{Doietal2010} and the k-correction, following the empirical k-corrections presented by \cite{OMilletal2011}.

\begin{figure*}
\centering
\includegraphics[scale=0.45]{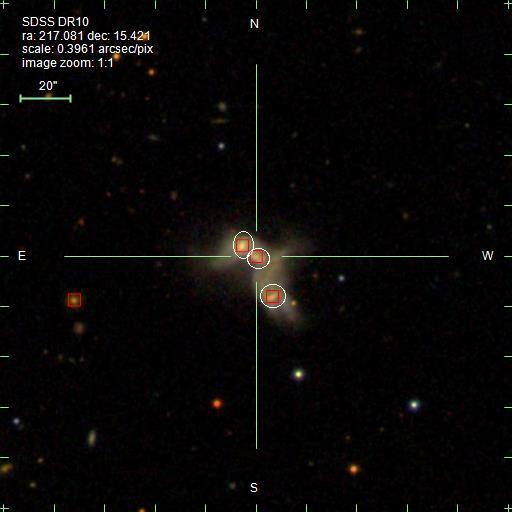}
\includegraphics[scale=0.45]{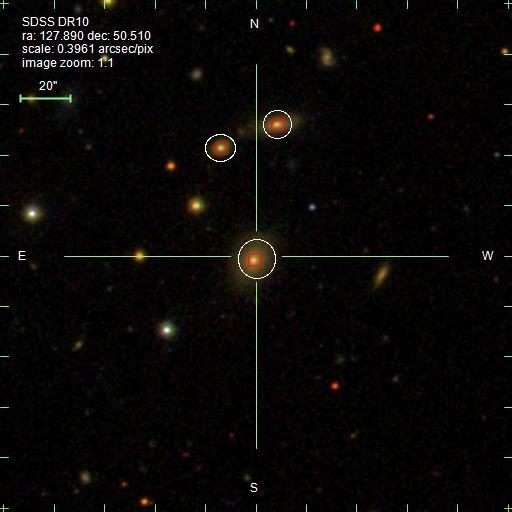}
\caption{Examples of the isolated galaxy triplets at $z=0.055$ (left panel) and $z=0.09$ (right panel). Figures extracted from \textit{http://skyserver.sdss.org/dr10/}. Galaxies with redshift determination are marked with a red box, and triplets members are indicated with white circles.}
\label{ex_trip}
\end{figure*}

\section{Triplet Identification}
\label{triplet_identification}

We have obtained a sample of galaxy triplet candidates in the volume-limited sample described in the previous section with the algorithm developed by \cite{OMilletal2012}. 

We consider as triplets the systems with three galaxies with spectroscopic observations that are close in projected separation  (r$_p \le $200 $h^{-1}kpc$) from the centre of the system (defined as the position of the brightest galaxy of the triplet) and have radial velocity differences with respect to the central galaxy $|\Delta V| \le$ 700 $km s^{-1}$. 

%
As our goal is to identify isolated galaxy triplets, it is necessary an isolation criteria that ensures that there are no significant external perturbations acting on the galaxy systems. For this reason, we impose the condition that there are no bright  ($M_r < -20.5+5 \log h_{70}$) neighbours within $0.5 h^{-1} Mpc$ from the triplet centre and with the same restriction on $|\Delta V|$ used to identify triplet members. \cite{Duplancicetal2015} used these criteria to isolate triplet galaxy systems. Moreover, $95 \%$ of the systems are at distances greater than 3$h^{-1} Mpc$ from their massive neighbours, implying that our isolation criterion is effective in selecting systems far away from groups and clusters of galaxies. We also impose an additional restriction that within 1$h^{-1} Mpc$, and with the same restriction on $|\Delta V|$ used to identify triplet members, there can be only one galaxy brighter than $M_r = -20.5+5 \log h_{70}$, and this cannot be brighter than the brightest galaxy in the triplet. Under these considerations, the final galaxy triplet sample contains 97 isolated systems with 291 spectroscopic galaxies in the redshift range 0.04$\le$z$\le$0.1.

Spectroscopic incompleteness has to be taken into account in the triplet identification. Due to observational issues \citep{Straussetal2002}, not all objects classified as galaxies and with $r<17.77$ in the SDSS survey, are spectroscopically observed. As a consequence, some of the systems previously identified may not be truly isolated triplets, but quartets or more complex galaxy configurations. In order to quantify this effect on the triplet catalogue, we used photometric redshifts to identify potential photometric members of the systems. The procedure adopted for this purpose is fully described in Appendix \ref{appendix_spec_completeness}. From this analysis, we discarded 14 systems from our triplet catalogue, because they presented potential photometric members that would break our isolation criteria.    

Additionally, after the spectral analysis three other triplets were excluded from our sample because they included galaxies classified as Seyfert type I and for them some stellar population properties (mainly stellar mass) could not be determined (see section \ref{spectral_analysis}). In summary, our final sample comprises 80 galaxy systems (or 240 galaxies) with spectral synthesis and emission line features determined. 

The high degree of isolation of the triplets can be appreciated in Figure \ref{ex_trip}, showing two galaxy triplets at $z=0.055$ and  $z=0.090$. Galaxies with redshift determination are marked with a red box, and triplets members are indicated with white circles. In the figure on the left, we can see a galaxy with spectroscopy, but its redshift is $z=0.269$ which leaves it out of our isolation criteria.

\section{Spectral Analysis}
\label{spectral_analysis}

\subsection{Empirical Spectral Synthesis}

The SDSS spectra were downloaded from the SDSS-DR10 database, corrected by galactic extinction using the \cite{Cardellietal1989} extinction law and transformed to the rest-frame. Each spectrum was then analyzed with the STARLIGHT spectral synthesis code \citep{CidFernandesetal2005}. A set of simple stellar populations from the library of \cite{BC03} with 150 elements (25 ages and 5 metallicities) was used, with ages between 10$^6$ and 18$\times$10$^9$ yr and metallicities between Z/Z$_{\odot}=$0.005 and 2.5. In order to carry out the spectral fitting, the STARLIGHT code uses a spectral mask to exclude regions around the emission lines and, consequently, only the stellar continuum is fitted. Representative stellar population features were defined using the resulting spectral synthesis light vector ($x_i$). The stellar population light-weighted age associated to a spectrum is defined as

\begin{equation}
<\log (t)>_L = \sum_i^{N_{SSP}} x_i \log (t_i)
\end{equation}
where the parameters $x_i$ and $\log (t_i)$ represent the light contribution at 4020\AA\ and the age of the i-th SSP element of the spectral library, respectively. It is important to mention that all weighted ages are truncated for values lower than the age of the Universe ($t\le13.5\times10^9$ years) in order to get ages consistent with the cosmology adopted. The SSPs older than the Universe are included in the spectral base because they can provide a better spectral fitting, specially of early-type galaxies. 

Stellar masses are calculated by taking into account aperture effects \citep{Straussetal2002}. The SDSS fiber has a 3'' diameter in the sky plane and then cannot cover all the extension of most galaxies. The fiber flux and consequently the stellar masses derived from these fiber spectra are consequently underestimated. To correct this effect, a procedure introduced by \cite{CidFernandesetal2005} is adopted. This correction compensates for the light which is lost by the fiber by multiplying the stellar mass inside the fiber by the factor $10^{0.4(z_{fiber} - z_{model})}$, where $z_{fiber}$ is the magnitude inside the fiber and $z_{model}$ is the \textit{ModelMag}, both at the $z$-band. This photometric band is chosen because the old stellar populations, which are usually responsible for most of the stellar mass, present their main emission at this band. This correction assumes that the mass-to-light ratio inside and outside the region covered by the fiber is the same. 

\subsection{Emission lines measures}

With the goal of investigating the emission lines properties of galaxies in triplets, the equivalent widths (EW) and fluxes of \Oiii, \Hb, \Ha\ and \Nii\ lines were measured by subtracting the model spectrum from the observed spectrum and then by fitting Gaussian profiles to these emission lines \cite[more details can be found in][]{Mateusetal2006,CidFernandesetal2010}. We consider as reliable measurements those emission lines with signal-to-noise greater than 3 and a fraction of bad pixels in the emission line region lower than 25\%. A fraction of galaxies (61 objects) could not fill these conditions and a visual inspection of their spectra was carried out. This inspection has indicated that 58 objects are passive galaxies, with no significant emission lines and predominantly old stellar populations. Three galaxies were classified as Seyfert I, with strong and broad emission lines. Passive and Seyfert I galaxies are not plotted in the WHAN diagram (see Figure \ref{WHAN_fraction}) either due to the absence of emission lines or because these emission lines could not be fitted by Gaussian profiles. 

The emission line analysis was done with a diagnostic diagram proposed by \cite{CidFernandesetal2011}. In this diagram (WHAN), the identification of the ionization mechanism allows the classification of galaxies as star-forming, AGNs (mostly Seyfert II and LINERS), passive (galaxies without significant emission lines) or retired galaxies (objects whose emission is attributed to advanced stages of stellar evolution). This analysis can be done by using the flux ratio  \nii/\Ha\ and the equivalent width of \Ha. In this emission line representation, galaxies with different ionization mechanisms occupy distinct loci. 

\section{General configuration of galaxy triplets}
\label{general_configuration}

The stellar mass in a galaxy triplet is an important quantity for an overall characterization of the system. It is calculated by summing the stellar mass obtained for each member galaxy (one of the outputs of the STARLIGHT synthesis). In the analysis below, we consider three stellar mass bins, divided by the 33.3\% and 66.6\% percentiles of the stellar mass distribution, or the intervals log(M$_*$)$<$11.52, 11.52$<$log(M$_*$)$<$11.72 and log(M$_*$)$>$11.72 (in solar units), respectively. In this section, we define the luminosity extent of triplets and verify how it correlates with triplet parameters such as total stellar mass and luminosity and mean age of the brightest galaxy. In the second part of this section, we use again STARLIGHT outputs to investigate the stellar mass assembly of the systems. Finally, we discuss the relation between mean stellar age and triplet stellar mass for the galaxies in our sample. 

\subsection{The luminosity extent of triplets}

To analyze each triplet, it is useful to consider the luminosity hierarchy of its members. We have classified each galaxy in a triplet as the brightest, the intermediate or the faintest galaxy of the system, accordingly to its $r$-band luminosity. This classification is relative to each system, i.e., the brightest object in a triplet may be fainter than the faintest member of another triplet. 

The relative luminosity extent of a system is defined as the ratio between the $r$-band luminosities of its brightest and faintest galaxies, $L_{3,1} = L_{r,faintest} / L_{r,brightest}$. We have also considered a second hierarchical parameter, for measuring the dominance of the brightest galaxy and defined as the fraction of the brightest galaxy luminosity with respect to the triplet luminosity: $f_L = L_{r,brightest}/L_{r,triplet}$. Since we noticed that both parameters are strongly correlated (the Spearman's correlation coefficient is -0.925, with the probability of the null hypothesis of absence of correlation $P(H_0)<10^{-3}$), our results are shown only for the parameter $L_{3,1}$. Our results would be similar if we had used $f_L$.

\begin{figure}
\includegraphics[scale=0.3]{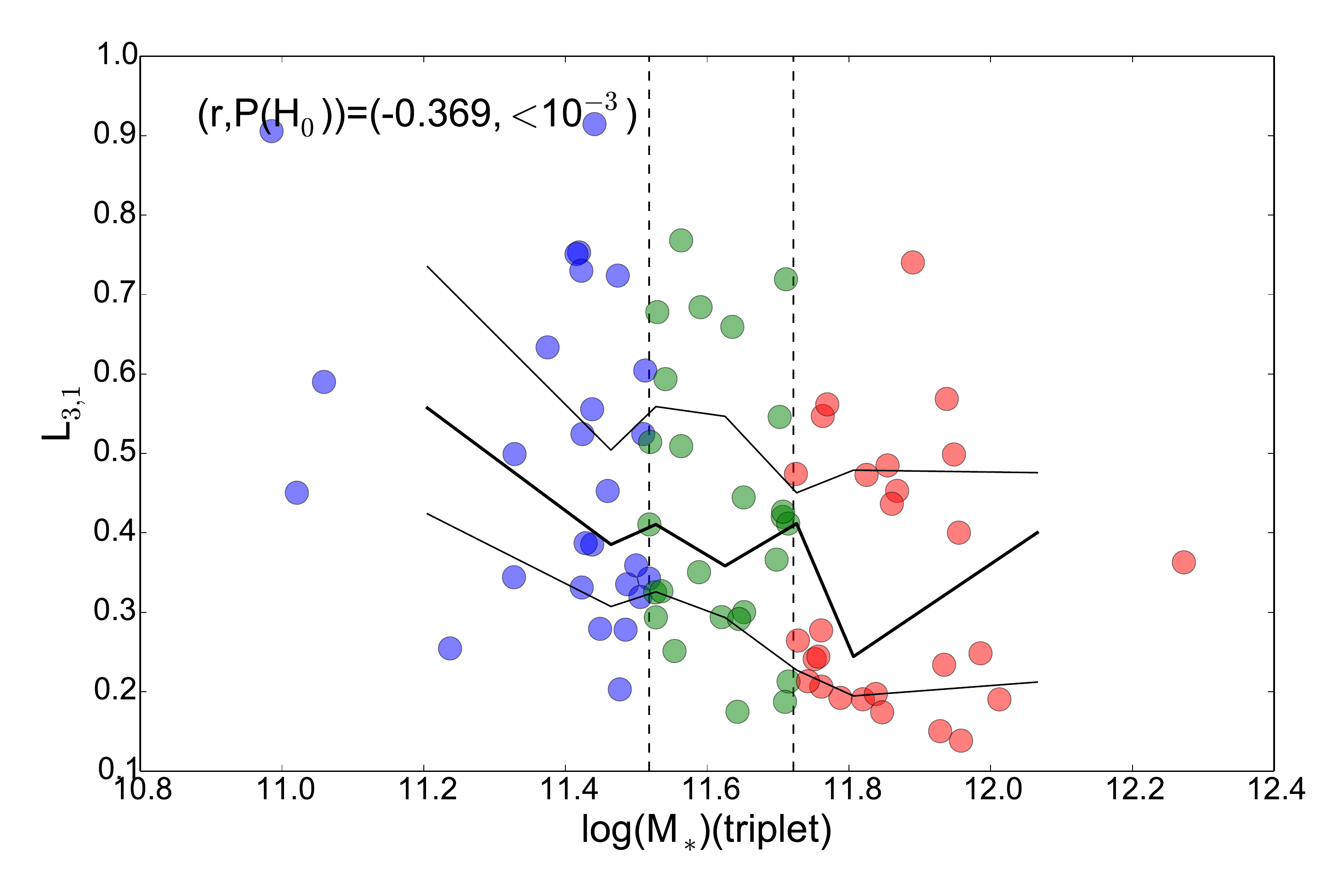}
\includegraphics[scale=0.3]{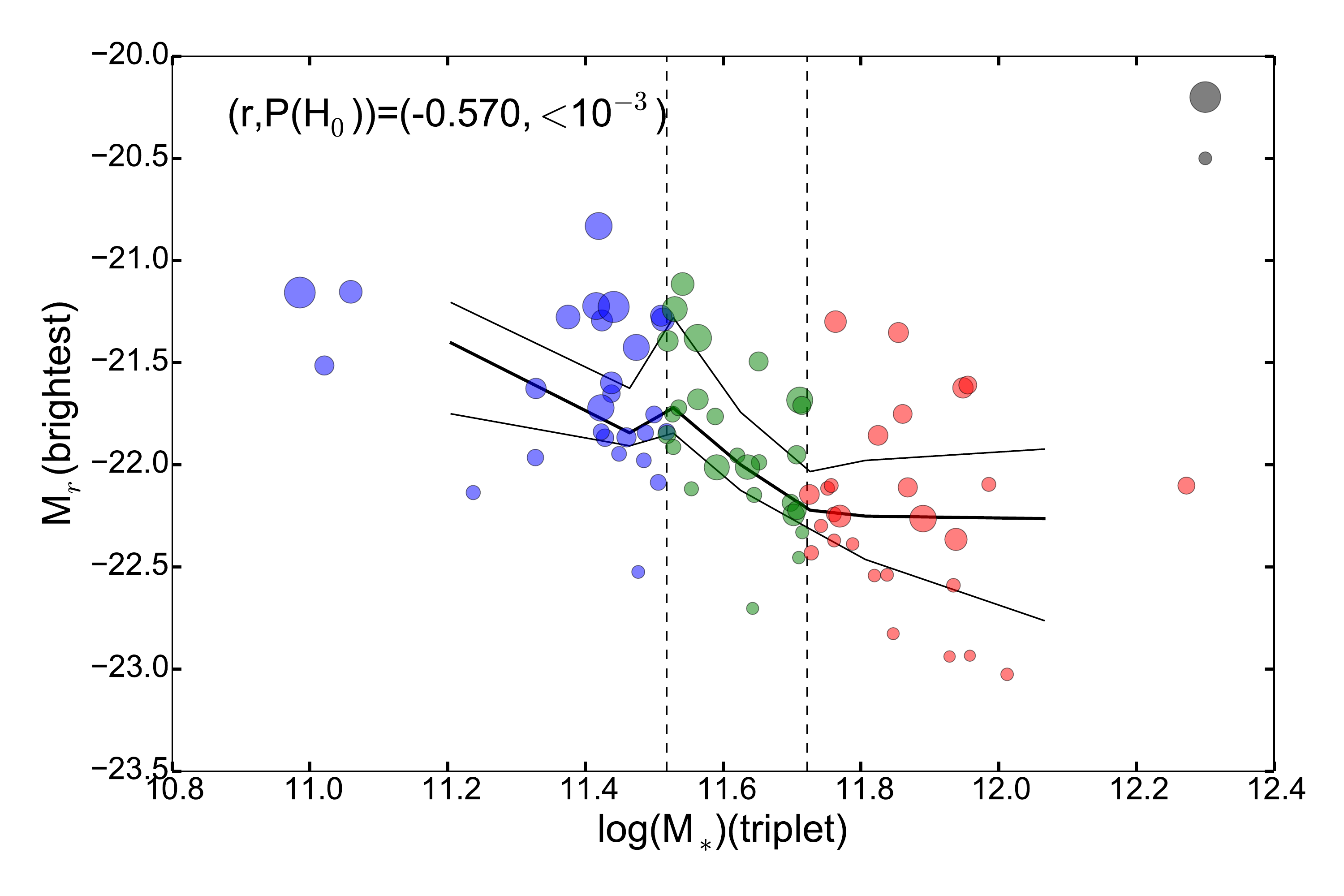}
\includegraphics[scale=0.3]{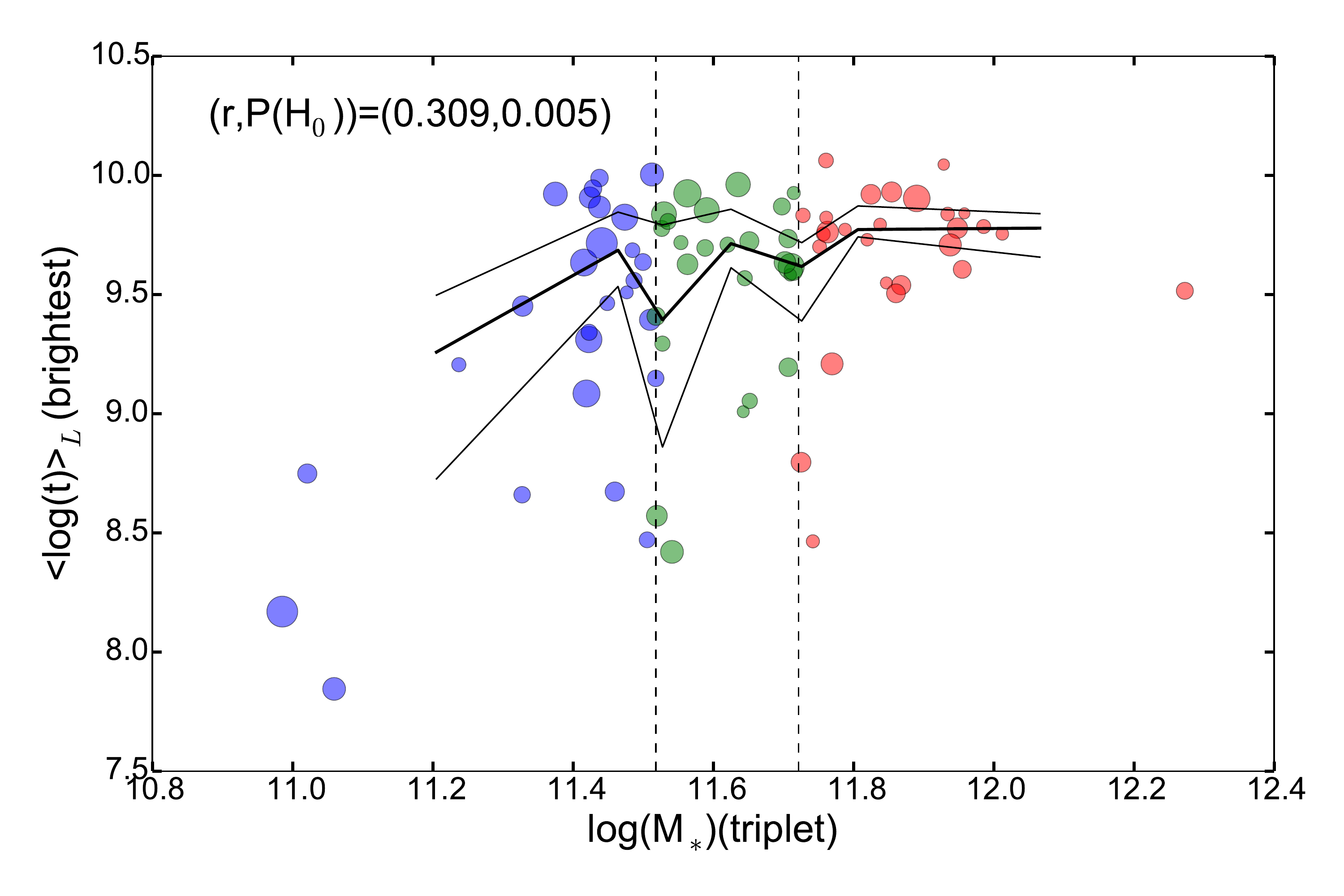}
\caption{Upper panel: The triplet stellar mass and the hierarchical parameter $L_{3,1}$. Middle panel: The stellar mass of the triplet and the $r$-band absolute magnitude of the brightest galaxy of each system; the small and large grey circles at the upper-right part of the panel represent $L_{3,1}$ equal to 0.2 and 0.9, respectively. Lower panel: The stellar mass of the triplet and the luminisity-weighted age of the brightest galaxy. Dashed vertical lines represent the stellar mass bins limits (see Section \ref{general_configuration}). The continuous lines connect the medians and quartiles of each bin. In the middle and lower panels, the circle size is proportional to the parameter $L_{3,1}$.}
\label{logM_L31}
\end{figure}

Firstly, an anti-correlation between triplet stellar mass and the parameter $L_{3,1}$ is found, i.e., $(r,P(H_0))=(-0.369,<10^{-3})$, as shown at the top of Figure \ref{logM_L31}. This result suggests that higher mass triplets tend to present lower values of $L_{3,1}$, i.e., the higher the dominance of the brightest galaxy is over the other members. On the other hand, low stellar mass triplets have higher values of $L_{3,1}$, i.e., the members have similar luminosities. The median value of $L_{3,1}$ for triplets in the most massive bin is 0.26, similar to the value adopted by \cite{Lambasetal2012} to separate between major and minor galaxy pairs. Some of the triplets in the most massive bin present extreme values such as $L_{3,1}\sim$0.2 or less, i.e., the luminosity ratio between the brightest and faintest galaxies is roughly five or more. It is important to highlight that this luminosity ratio represents a difference lower than 2 in the magnitude scale, which is the limit adopted by several authors \cite[e.g.][]{SalesLambdas2005} to select galaxy satellites. 

The middle panel of Figure \ref{logM_L31} shows the $r$-band absolute magnitude of the brightest galaxy ($M_r$) as a function of the triplet stellar mass. In this figure circles represent triplets and their sizes are proportional to the parameter $L_{3,1}$. An anti-correlation is found $(r,P(H_0))=(-0.570,<10^{-3})$, showing that the brightest triplet galaxies are more luminous (or more massive) in more massive triplets. This configuration is expected in the hierarchical scenario, acting in systems simpler than groups and clusters. Moreover, one can notice that as the absolute magnitude of the brightest galaxy increases, the dominance parameter $L_{3,1}$ decreases, for a certain triplet mass value.  

The lower panel of Figure \ref{logM_L31} shows an initial stellar population analysis of our triplet sample. From this figure it can be appreciated a trend between the age weighted by the light of the brightest member ($<log(t)>_L$) and the stellar mass of the triplet, with a Spearman correlation of $(r,P(H_0))=(0.309,0.005)$. Brightest galaxies with older stellar populations galaxies are common in massive triplets, but less prevalent in the lower mass bin. If the same analysis is done using the faintest galaxy of the triplet, no significant correlation is found. This suggests that the triplet properties are mainly correlated to the brightest galaxy. Fainter members present a secondary role on triplets properties.  

We now explore the spectral properties of triplet galaxy systems with other techniques, such as the mass assembly history obtained from spectral synthesis analysis and the emission line diagnostic diagrams.

\subsection{The stellar mass assembly}
\label{mass_assembly}

In this section, we consider how the mass assembly history of the triplets can provide a more detailed understanding of the evolution of their galaxies. We first discuss how the empirical spectral synthesis approach adopted by the STARLIGHT code allow us to retrieve these histories and, after, we present results for the mass assembly of galaxies as a function of their place in the luminosity hierarchy of the triplets.

One of the spectral synthesis results is the mass vector ($\mu_{ini}(t_i)$), which gives the fraction of stellar mass in SSPs of age $t_i$ (including the mass returned to the ISM). Following  \cite{Asarietal2007}, we determine the fraction of the current mass in a galaxy at a certain look-back time $t_*$ as 
\begin{equation}
\centering
\eta(t_*) = \sum_i^{t<t_*} \mu_{ini}(t_i).
\end{equation}
This is a cumulative function that grows from 0 to 1 (for $t_*$ equal to the age of the Universe), starting at the oldest SSP in the spectral base and tracking what fraction of the stellar mass was due to stars formed up to a given look-back time. We have adopted here a procedure proposed by  \cite{Asarietal2007}, where the mass vector $\mu_{ini}(t_i)$ is resampled in time, with $\Delta log(t)$=0.1 dex, and then it is smoothed using a Gaussian filter with standard deviation equal to $\Delta log(t)$, to avoid discontinuities associated with the discrete set of ages in the base.

\begin{figure}
\centering
\includegraphics[scale=0.4]{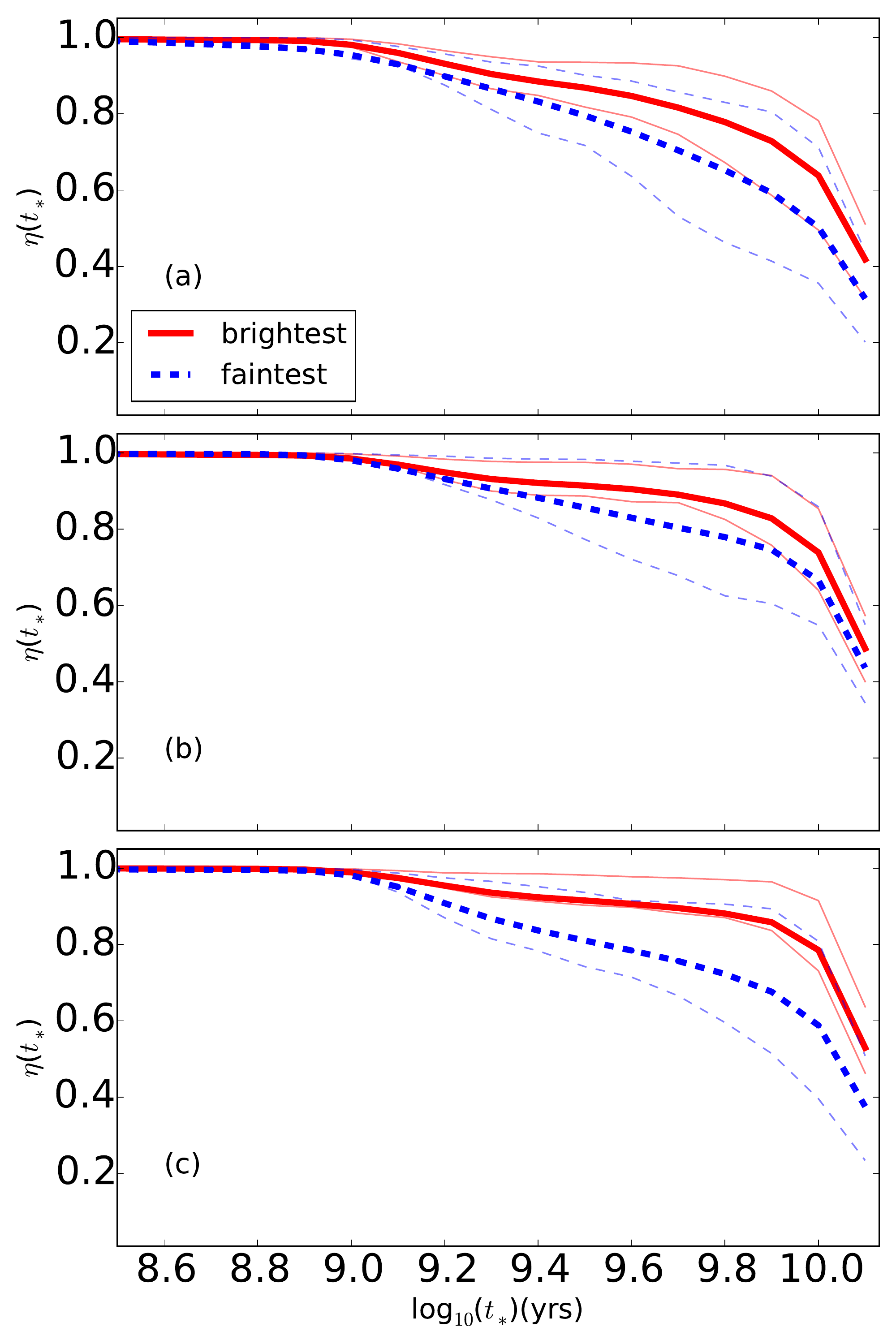}
\caption{Mean mass assembly of galaxies classified according to their triplet mass bin, i.e., low triplet mass (a), intermediate triplet mass (b) and high triplet mass (c). The brightest and faintest galaxies in triplets are represented as thick continuous and dashed lines, respectively. Lighter lines represent 25\% and 75\% percentiles of dispersion.}
\label{mass_assembly_brightest_faintest}
\end{figure}

\begin{figure*}
\centering
\includegraphics[scale=0.35]{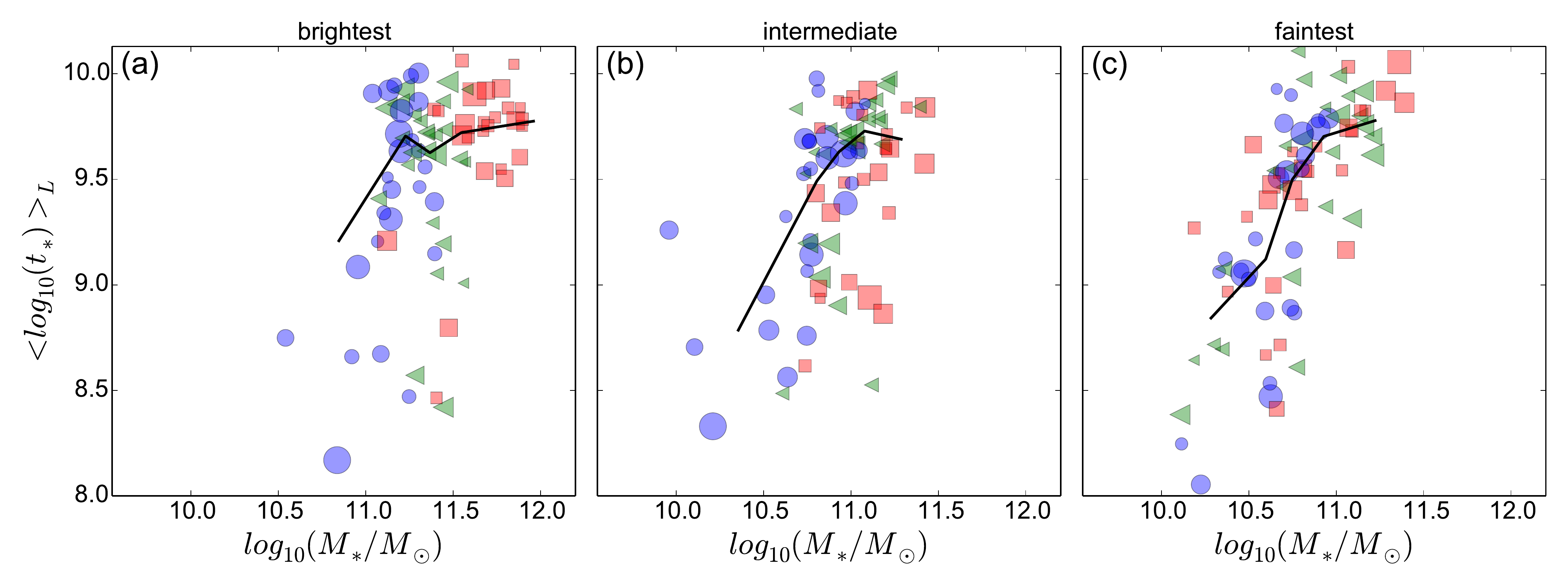}
\caption{The age - stellar mass relation for triplet galaxies considering different hierarchical classes and triplet stellar mass bins. The panels (a), (b) and (c) show this relation for galaxies classified as brightest, intermediate and faintest ones, respectively. Circles, triangles and squares represent galaxies in low, intermediate and high stellar mass triplets, respectively. Their sizes are proportional to the L$_{3,1}$ parameter and solid lines represent the median trend.}
\label{Mcor_logt_brightest_faintest}
\end{figure*}

This quantity shows how fast a certain galaxy has formed its stars, being different for distinct Hubble types: while early-type galaxies tend to form the bulk of their stars a long time ago, the stellar mass assembly of star-forming galaxies is slower. \cite{Asarietal2007} have shown that the star formation history of star-forming galaxies also vary systematically along the star-forming sequence of the BPT diagram (i.e., along the right wing in the \Oiii/\Hb\ versus \Nii/\Ha\ plane), with low-metallicity systems evolving at a slower pace and with a current large star formation rate than their high-metallicity counterparts, as expected in the downsizing scenario \citep{Cowieetal1996}.

Figure \ref{mass_assembly_brightest_faintest} shows the mean stellar mass assembly of galaxies in different luminosity hierarchical classes and in different mass bins. The thick lines correspond to the mean behaviour of each class and the light lines represent the 25\% and 75\% percentiles of the distribution of $\eta$ for objects of each class. These curves are shown up to log(t)=8.5 (for $t$ in years) since the bulk of the stellar mass is formed earlier than this lookback time. This figure shows that, despite the large scatter, the brightest triplet galaxies tend to form stars more quickly than the faintest triplet galaxies, again consistent with the expectations of the downsizing effect \citep{Cowieetal1996}. Although not shown in this figure, intermediate galaxies in triplets tend to have a stellar mass assembly behaviour intermediate between the two cases above.

Comparing now the mean mass assembly of galaxies with the same position in the luminosity hierarchy but in different triplet mass bins, we notice that the most massive triplets harbour galaxies which have formed their stars faster in the past than galaxies in less massive triplets. This effect is significant for the brightest galaxy in a triplet but less conspicuous for the intermediate and  faintest galaxies. The stellar mass range for the intermediate mass bin is quite narrow when compared to the others and can present objects with similar properties to the low mass bin. 

\subsection{The age - stellar mass relation}

The relation between the mean stellar population age and the stellar mass of galaxies illustrates how galaxies evolve. Figure \ref{Mcor_logt_brightest_faintest} shows this relation for galaxies within different hierarchical and triplet mass classes. 

The age - stellar mass relation of galaxies present different slopes according to their hierarchical classes. The brightest galaxies in triplets present a distinct slope in this relation when compared to the other hierarchical classes. These galaxies are mostly clustered at higher stellar masses and older stellar population ages, producing a low slope relation when compared to the other classes. In addition, the brightest galaxies from high and low mass triplets mostly occupy the highest and lowest stellar mass region of this relation, respectively. There is again a relation between the properties of the brightest galaxy and the triplet properties. This result is not noticed for the intermediate and the faintest galaxies of the triplet. In this case, the intermediate and the faintest galaxies are not clustered in any region of the age - stellar mass plane according to their stellar mass triplet class.    

These results suggest that the evolution of faintest and intermediate galaxies are quite similar but the brightest triplet galaxies have evolved in a different way.

\section{Emission lines analysis}
\label{elines}

\subsection{The WHAN diagram}

The WHAN diagram allows the investigation of the ionization mechanism in galaxies, responsible for the emission lines. In the EW(\Ha) - \Nii/\Ha\ plane, galaxies can be classified as star-forming (actively forming stars), retired/passive (without or with negligible star formation)  and AGN (either weak or strong). Here we search for correlations of WHAN classes with the luminosity hierarchy and with the global properties of the triplets.

\begin{figure*}
\centering
\includegraphics[scale=0.4]{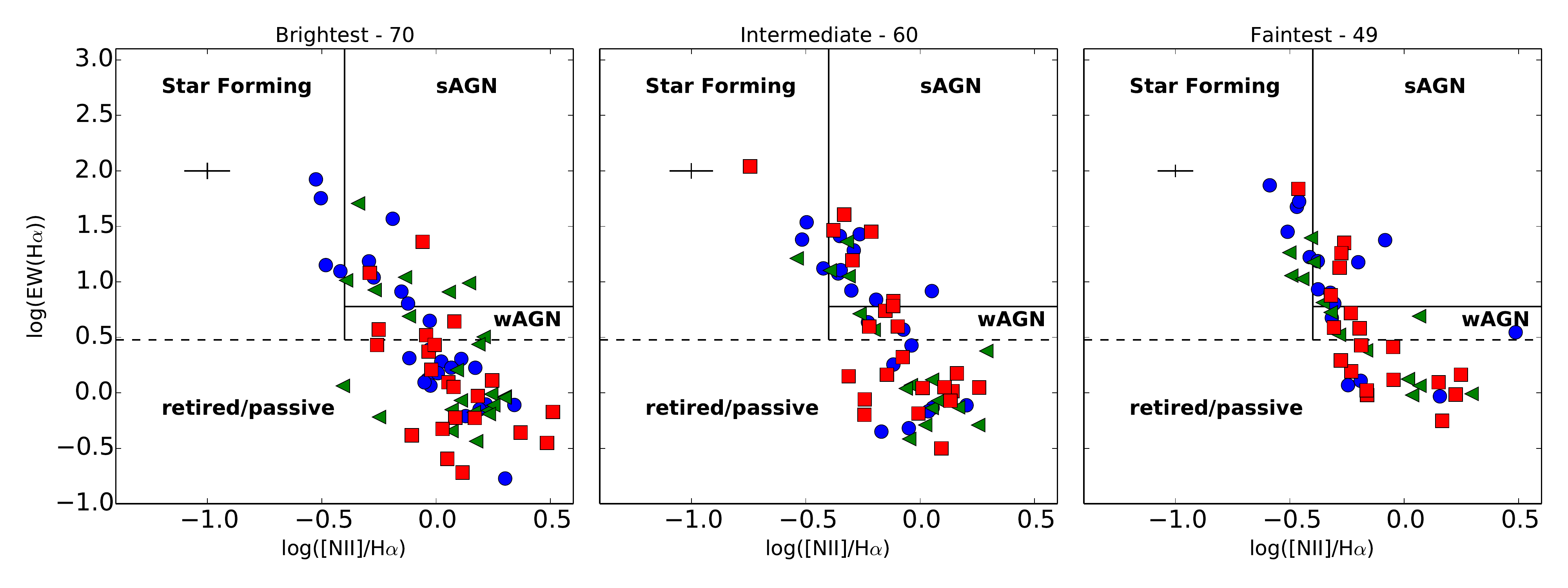}
\includegraphics[scale=0.4]{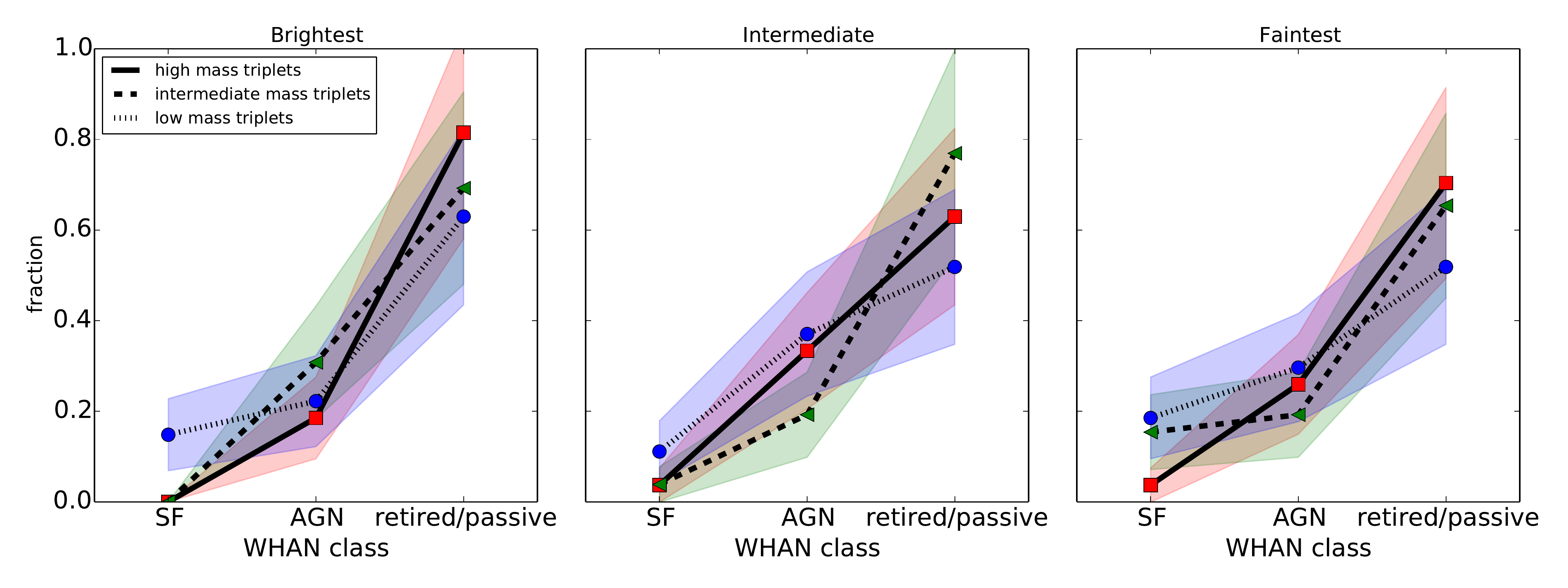}
\caption{Upper panel: The WHAN diagram for galaxies in triplets according to their luminosity hierarchy: brightest (left), intermediate (center) and the faintest (right). The symbols represent galaxies from different triplet mass bins: red squares- high mass triplets, green triangles- intermediate mass triplets and blue circles- low mass triplets. Mean errors are represented at the upper-left region of each panel. The number of galaxies plotted in each panel is shown on the top. Lower panel: The fraction of galaxies classified as star-forming (SF), AGN (strong and weak AGNs) and passive/retired galaxies, according to their hierarchy and triplet mass bins. Shaded areas represent 1$\sigma$ uncertainties.}
\label{WHAN_fraction}
\end{figure*}

The upper panels of Figure \ref{WHAN_fraction} show the WHAN diagram for triplet galaxies according to their luminosity hierarchical class and triplet mass bin. Some galaxies do not appear in this figure, mostly due to absence of emission lines or for difficulties in measuring the lines (see Section \ref{spectral_analysis}). For this reason, the fraction of passive (mostly lineless spectra) in the figure is underestimated. From the panels in this figure, it can be noticed that the majority of triplet galaxies are located in AGNs, passive and retired galaxies loci, and only a few triplets present galaxies classified as star-forming. This result is valid for galaxies classified as the brightest, intermediate and the faintest in the triplets. The distributions of the galaxies in this diagram, regardless of the triplet mass bin, are different according to their hierarchical class in the triplet. 

We have evaluated the fraction of the galaxies in each WHAN class, considering also triplet masses and galaxy luminosity hierarchical classes, as shown in the lower panels of Figure \ref{WHAN_fraction}. In general, we can notice an increase in the fractions from star-forming galaxies to AGN and to passive and retired galaxies. The large fraction of passive and retired galaxies is clear for all hierarchical classes and triplet mass bins. The fraction of star-forming galaxies is the lowest, being at most $\sim$20\% for low mass triplets. Part of this result may be due to a selection effect. As a consequence of the volume-limited galaxy sample, low stellar mass galaxies are left out from the analysis, what can affect  these fractions.

In the triplet low mass bin, there is a systematically higher fraction of star-forming galaxies compared to the high mass bin. Star-forming galaxies are not found among the brightest galaxies in massive triplets and are not frequent also among intermediate mass triplets. Comparing the fractions in the highest and lowest mass bins, the main trends are: similar fractions of strong and weak AGNs, higher fraction of star-forming galaxies in low mass triplets and higher fraction of retired and passive galaxies in high mass triplets. This result is more prominent comparing the fractions for brightest and faintest galaxies. 

These results suggest that the environment in the low mass galaxy systems is not as efficient to halt star formation as it is in massive systems. They also show the importance of the triplet mass on the evolution of the stellar populations of the galaxies in the system, for all luminosity hierarchical classes.

\begin{figure*}
\centering
\includegraphics[scale=0.38]{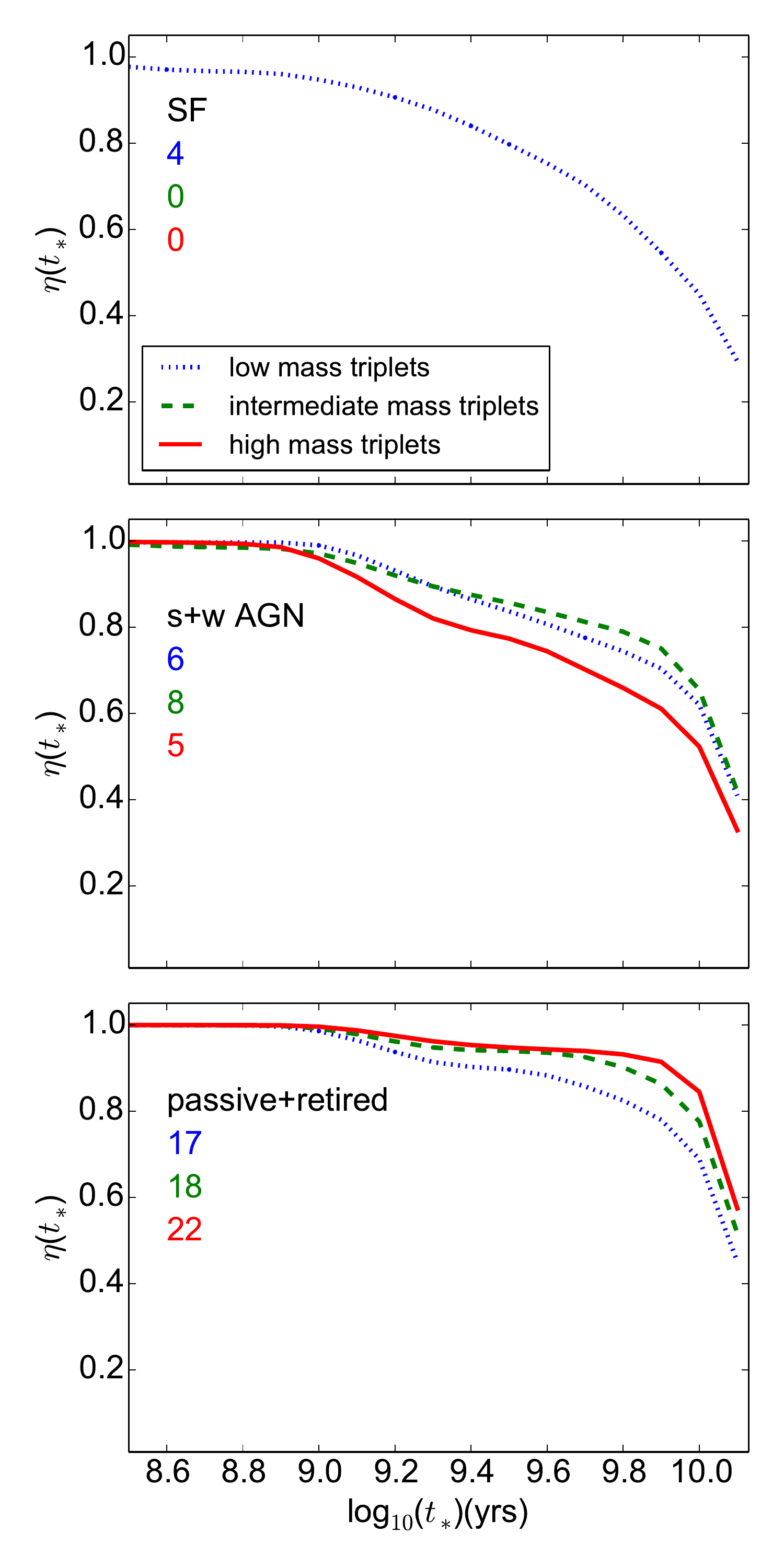}
\includegraphics[scale=0.38]{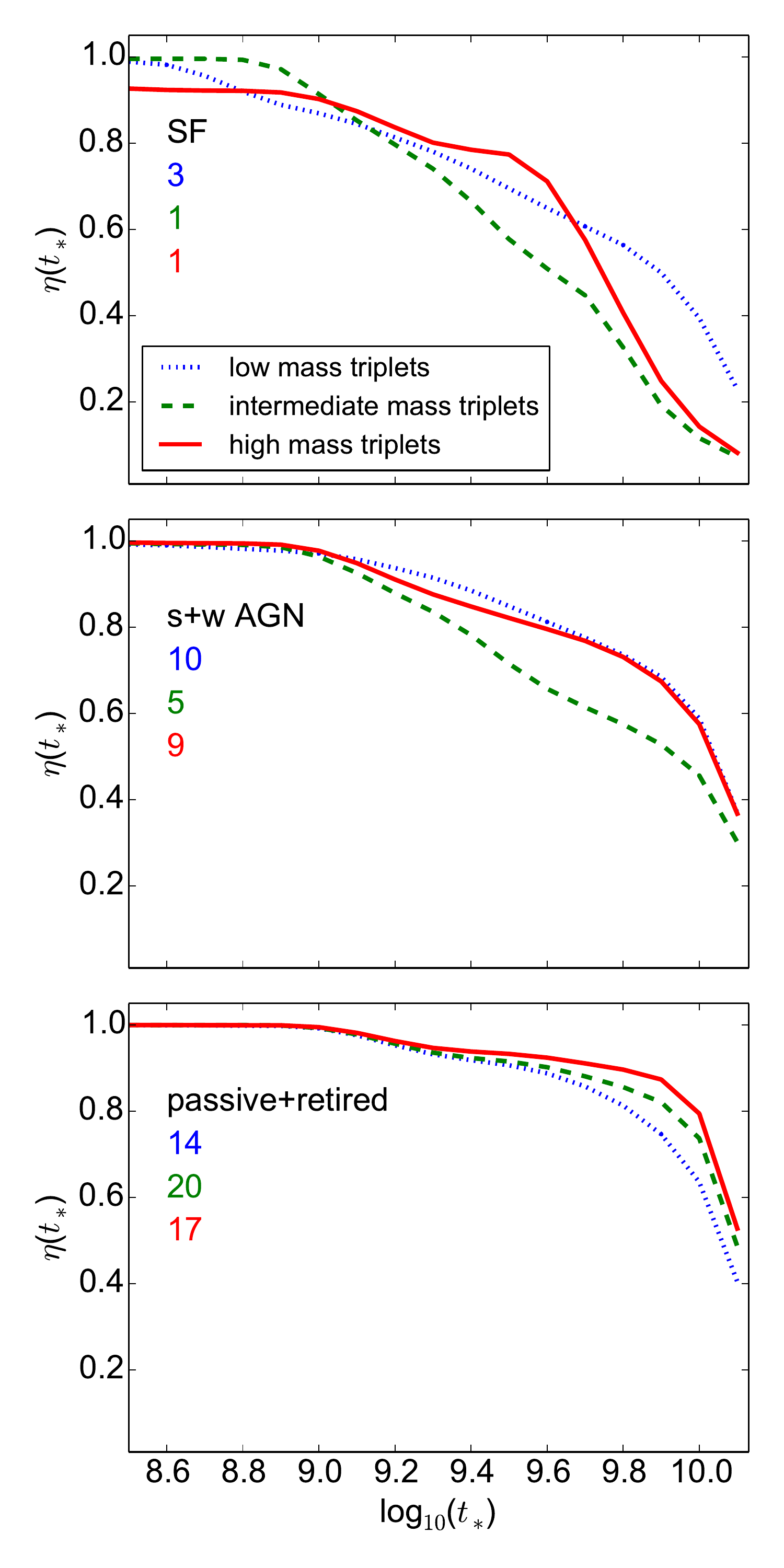}
\includegraphics[scale=0.38]{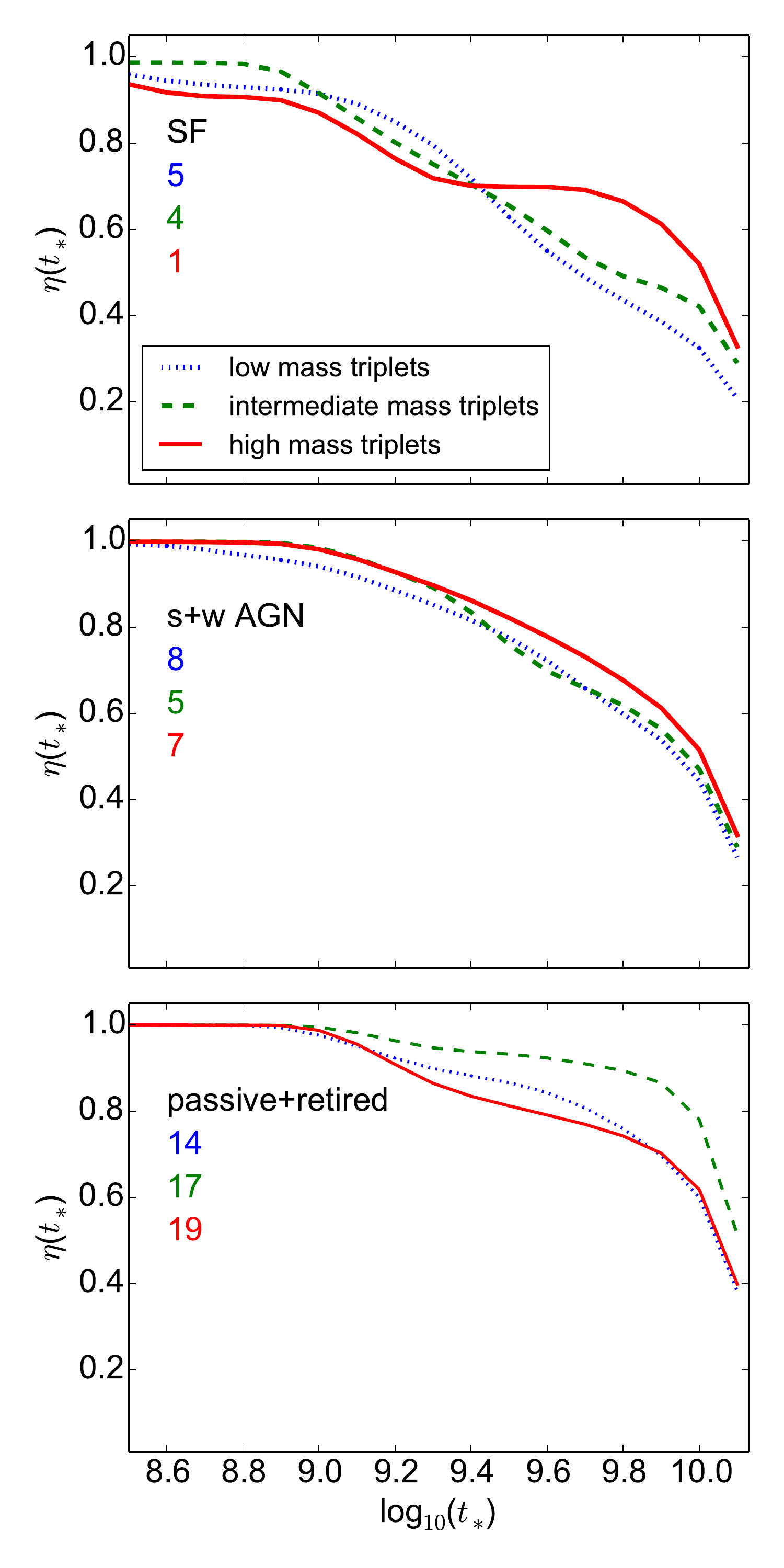}

\caption{Mean mass assembly of galaxies in triplets, classified as star-forming (SF), strong and weak AGNs (s+w AGN) and passive and retired galaxies, for different luminosity hierarchy and triplet mass bins. Left: Brightest galaxies in triplets. Center: Intermediate galaxies in triplets. Right: Faintest galaxies in triplets. Different curve types represent distinct mass bins. The number of galaxies used to calculate the mean mass assembly history for each WHAN and hierarchical classes is shown in all panels.}
\label{mass_assembly_WHAN_bright_faint}
\end{figure*}

\subsection{The mass assembly of WHAN classes}

Now we discuss the stellar mass assembly of galaxies in the several WHAN classes. Figure \ref{mass_assembly_WHAN_bright_faint} shows the mass assembly history curves for galaxies in different luminosity hierarchical classes and classified according to the WHAN diagram. The 25\% and 75\% percentiles of the mass assembly history are omitted for the sake of clarity in this figure. 

The mean mass assembly history of galaxies classified as star-forming are basically the same in all mass bins and hierarchical classes. The same behaviour can be noticed for weak and strong AGNs. 

Galaxies classified as passive and retired, however, present distinct assembly histories for different luminosity hierarchical classes, mainly when we compare the brightest and faintest galaxies. Passive or retired galaxies, in the most massive triplets, have formed the majority of their stars in epochs earlier than the brightest ones in less massive triplets. Since the most of brightest galaxies are classified as passive and retired, then a similar behavior for the mass assembly history is expected for this WHAN class, as shown in Figure \ref{mass_assembly_brightest_faintest}. Again, the mass assembly history of massive galaxies depends on the triplet mass; this trend, however, is not observed in the other mass bins nor for the intermediate and faintest triplet galaxies.

\subsection{The combination of WHAN classes according to triplet properties}

Here we investigate the combinations of galaxies in each triplet, according to their WHAN classes. A census of all combinations of star-forming (1), strong + weak AGNs (2) and passive and retired galaxies (3) can reveal how WHAN classes are distributed in different triplet stellar mass bins. For instance, a combination of three star-forming galaxies is expressed as 111, whereas, in the same notation, two galaxies classified as AGNs and one as passive or retired is named 223. 

\begin{figure*}
\centering
\includegraphics[scale=0.4]{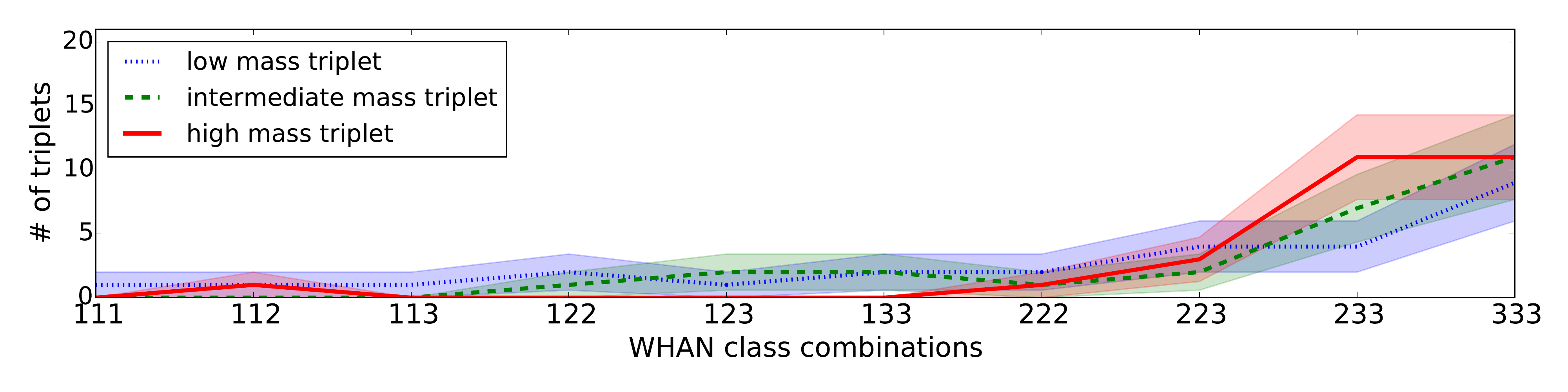}
\caption{Fraction of triplets with several combinations of WHAN classes (star-forming (1), s+w AGN (2), passive/retired(3)), for different triplet stellar mass bins. The fraction of passive+retired galaxies increases to the right while the star-forming fraction decreases. Shaded areas represent Poisson errors.}
\label{frac_combination_WHAN}
\end{figure*}

Figure \ref{frac_combination_WHAN} shows the fraction of galaxy combinations according to WHAN classes for each mass bin. Only three triplets contain the majority of star-forming galaxies (classified as 111, 112 and 113), mostly in the low mass  bin. This result is in agreement with \cite{Duplancicetal2013}, who found that blue triplets are located in the less massive tail of the total stellar mass distribution and present an efficient star formation activity when compared to compact groups, in the same stellar mass bin. As the fractions of AGNs and passive galaxies increases, the number of triplets with a dominance of passive and retired galaxies (mainly 233 and 333) increases. This is an indication that, within the framework of the WHAN classification and for the mass interval considered here, the bulk of the triplets are dominated by passive and retired galaxies and, in a smaller fraction, AGNs. The fractions of triplets dominated by passive or retired galaxies (133, 233 and 333) are 55\%, 77\% and 81\% in the lowest, intermediate and highest mass bins, respectively.

\begin{figure}
\centering
\includegraphics[scale=0.29]{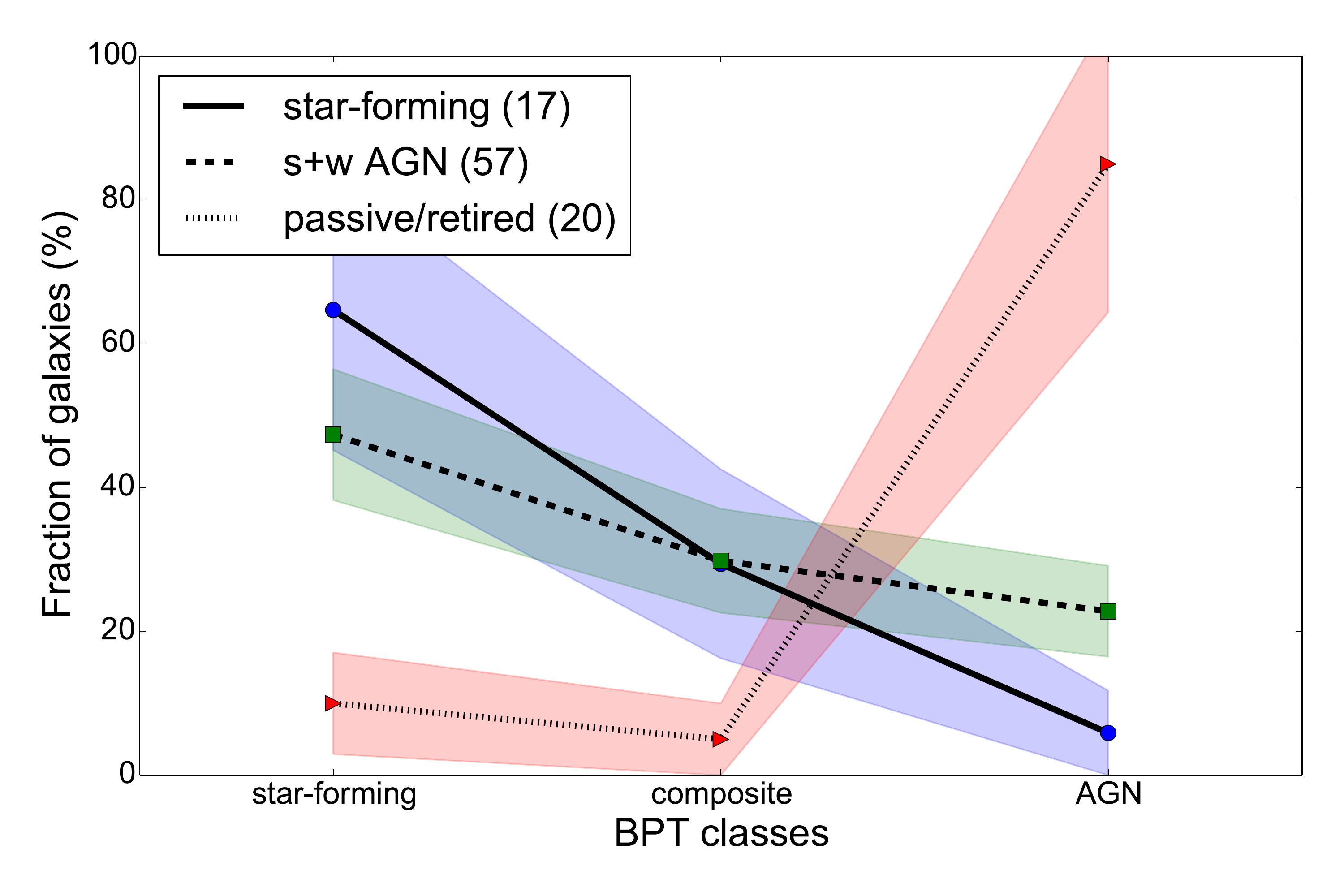}
\caption{Fraction of galaxies classified in the WHAN diagram as star-forming (continous line), strong and weak AGNs (dashed line) and passive + retired (dotted line), and their classifications according to the BPT diagram. Shaded areas represent Poisson errors. The number of galaxies in each class is also shown.}
\label{WHAN_BPT_classes}
\end{figure}

\subsection{Comparison with BPT diagram}

We can also make use of other diagnostic diagrams to classify galaxies according to their emission lines. The BPT diagram \citep{BPT81} is traditionally employed to evaluate the photoionization source of galaxies. Therefore, it is useful to compare our WHAN-based classes with the BPT diagram classification. 

While the WHAN diagram requires two emission lines (\Ha\ and \Nii) to classify a galaxy, the BPT diagram needs two more, namely \Hb\ and \Oiii. This diagram basically classifies galaxies into three classes: star-forming, composite and AGNs. The composite class is an intermediate class, containing star-forming and AGN contributions, and defined here as the region between two curves, theoretically and empirically defined by \cite{Kewleyetal2001} and \cite{Kauffmannetal2003}, respectively. Obviously, some galaxies can be classified in the WHAN diagram but not in BPT due to the signal-to-noise and bad pixels fraction restrictions (see Section \ref{spectral_analysis}). We found 94 galaxies with classification in both BPT and WHAN diagrams.  

The comparison between the classifications of these 94 galaxies in both diagrams is shown in Figure \ref{WHAN_BPT_classes}. Roughly 60\% of the star-forming galaxies in the WHAN diagram are in agreement with the BPT classification, the fractions being smaller for composite and AGNs classes. The objects classified as weak and strong AGNs in the WHAN diagram are mostly classified as star-forming and composite objects in the BPT diagram, and only $\sim$30\% of the galaxies are classified as AGNs. Passive and retired galaxies interestingly present more than 80\% of galaxies classified as AGNs in the BPT diagram. This noticeable divergence between the two classification schemes is explained by the fact that the BPT diagram cannot distinguish galaxies with strong or weak emission lines since it only uses emission line ratios. \citet{Stasinskaetal2015} discussed in details this issue. Passive and retired galaxies, which present small equivalent width of \Ha\ in the WHAN diagram, are located in the same locus of AGNs in the BPT diagram. Only the WHAN diagram can evaluate galaxies according to their emission line strengths and consequently distinguish galaxies classified as retired or passive from AGNs.

\section{Summary and Conclusions}
\label{summary_conclusions}

In this work, we have used the algorithm described in \cite{OMilletal2012} to identify isolated galaxy triplets in a volume-limited sample ($M_r<-20.5+5\log h_{70}$) from the Sloan Digital Sky Survey Data Release 10. An initial sample of 97 galaxy systems between 0.04$\le$z$\le$0.1 is compiled. The code STARLIGHT \citep{CidFernandesetal2005} was used to provide information about the stellar population and the mass assembly history of triplet member galaxies. From the fluxes and equivalent widths of emission lines, the diagnostic diagram WHAN was used to classify galaxies, according to their ionization mechanism, into star-forming, strong and weak AGNs, or passive and retired galaxies. Due to the identification of some galaxies classified as Seyferts I in our triplets and spectroscopic incompleteness, our initial galaxy triplet sample is reduced to 80 objects with stellar population properties and emission lines measured for all triplet members. 
 
Our main conclusions are:
\begin{itemize}

\item There is an anti-correlation between the triplet stellar mass and the luminosity hierarchical parameter $L_{3,1}$, indicating that the luminosity dominance of the brightest galaxies increases with the triplet mass. Correlations between the properties of the brightest galaxy, such as age and absolute magnitude at $r$-band and the stellar mass of the triplet are significant.

\item The mean mass assembly history of galaxies, separated according to the hierarchy in the triplet (brightest, intermediate and faintest) and triplet stellar mass (high, intermediate and low), shows that the mass assembly of brightest galaxies also depends on the triplet mass. Brightest galaxies in more massive triplets formed the bulk of their stars earlier than those in low mass triplets. This behaviour is not so clear for the other galaxies, those in the intermediate and faintest position in the luminosity hierarchy of these systems.   

\item The relation between the mean stellar population age and the stellar mass of galaxies in triplets presents similar behaviour for galaxies classified as faintest and intermediate, and a distinct slope for the brightest triplet galaxies. This result suggests that the evolution of the intermediate and faintest galaxies is similar but the brightest galaxies present a distinct evolution, essentially characterized by a faster mass assembly.

\item The WHAN diagram shows that most galaxies in our sample are classified as AGNs or passive and retired galaxies. The dominance of passive and retired galaxies in almost all galaxy luminosity classes and triplet mass bins is evident, as shown in Figure \ref{WHAN_fraction}. Less massive triplets show different fractions for the different hierarchical classes. The environment of triplets favours interactions between the galaxies they contain \citep[e.g.,][]{Hernandez-Toledoetal2011}, and we can expect that these interactions grow with triplet mass, leading to a faster evolution of these structures and, in particular, producing most of the stars fastly in the past and leading nowadays to a population dominated by passive and retired galaxies. In low mass triplets, the interaction between galaxies did not arrive to halt star formation yet, leading to a different population class mix. 

\item The mass assembly history of galaxies classified according to the WHAN diagram indicates that star-forming galaxies present similar formation histories, independently of the triplet stellar mass or galaxy position in the luminosity hierarchy. On the other side, passive and retired galaxies had a fast mass assembly history. As most of the brightest galaxies are passive and retired, this result confirms our previous findings considering only the position in the luminosity hierarchy. 

\item The combinations of WHAN classes for triplet member galaxies in different triplet stellar mass bins was also explored in this work. Star-forming galaxies are mainly in low mass triplets and only a few triplets present the majority of their members with this classification. Most triplets are dominated by passive and retired galaxies, presenting two or more members in this class. The number of systems with this specific configuration is higher in more massive triplets.   

\item A comparison between the BPT and WHAN classifications highlights some differences between classes obtained with these two approaches. The BPT diagram is arguable more efficient than the WHAN diagram for classification of star-forming galaxies (because the use of four emission lines carries more information than just two among them), but it lacks efficiency to distinguish between retired galaxies from AGNs and is insensitive to passive galaxies. 
\end{itemize} 
 
The results presented in this work indicate that the brightest triplet galaxy is driving the evolution of the system, as suggested by the strong correlations between properties as mass assembly and stellar population age with triplet properties. Intermediate and faintest objects have a secondary role. Triplets are dominated by passive and retired galaxies. This fraction is maximum in high mass triplets. Taken as a whole, these results are consistent with the environment, through galaxy interactions, playing a major role in triplet evolution. This is consistent with an analysis of triplets in mock catalogues, which indicates that these systems are dynamically evolved and probably have undergone merger events \citep{Duplancicetal2015}, favouring the formation of early-type galaxies, classified spectroscopically as passive or retired galaxies.

\section*{Acknowledgments}

MVCD thanks his scholarship from FAPESP (process number 2014/18632-6). 
LSJ acknowledges the support of FAPESP (2012/00800-4) and CNPq to his work.

This work was supported in part by the Consejo Nacional de Investigaciones Cient\'\i ficas y T\'ecnicas de la Rep\'ublica Argentina (CONICET), Secretar\'\i a de Ciencia y Tecnolog\'\i a de la Universidad de C\'ordoba and Secretar\'\i a de Ciencia y Tecnolog\'\i ca 
de la Universidad Nacional de San Juan.\\
Funding for SDSS-III has been provided by the Alfred P. Sloan Foundation, the Participating Institutions, the National Science Foundation, and the U.S. Department of Energy Office of Science. The SDSS-III web site is http://www.sdss3.org/. 
SDSS-III is managed by the Astrophysical Research Consortium for the Participating Institutions of the SDSS-III Collaboration including the University of Arizona, the Brazilian Participation Group, Brookhaven National Laboratory, Carnegie Mellon University, University of Florida, the French Participation Group, the German Participation Group, Harvard University, the Instituto de Astrofisica de Canarias, the Michigan State/Notre Dame/JINA Participation Group, Johns Hopkins University, Lawrence Berkeley National Laboratory, Max Planck Institute for Astrophysics, Max Planck Institute for Extraterrestrial Physics, New Mexico State University, New York University, Ohio State University, Pennsylvania State University, University of Portsmouth, Princeton University, the Spanish Participation Group, University of Tokyo, University of Utah, Vanderbilt University, University of Virginia, University of Washington, and Yale University.

This work has made use of the computing facilities of the Laboratory of Astroinformatics (IAG/USP, NAT/Unicsul), whose purchase was made possible by the Brazilian agency FAPESP (grant 2009/54006-4) and the INCT-A.

\nocite{*}

\bibliographystyle{mn2e}


\appendix
\section{The Influence of the Spectroscopic Incompleteness on Triplet Identification}
\label{appendix_spec_completeness}

The spectroscopic incompleteness represents the main source of fake triplets in our sample. The fiber collision effect and the finite number of fibers of the spectrograph have as a consequence that not all galaxies in a magnitude-limited photometric sample are spectroscopically observed \citep{Straussetal2002}. Consequently, due to this effect, some of the systems identified from a spectroscopic sample as a galaxy triplet can, actually, be quartets or even systems with larger multiplicity. 
  
We employ photometric redshifts to investigate this incompleteness. The SDSS database provides photometric redshifts (table \textit{Photoz}) for objects classified as galaxies that can be used to evaluate more precisely these spectroscopic candidates to triplet member. Photometric redshifts, however,  contain larger uncertainties and may be biased compared to spectroscopic redshifts. In order to investigate this effect, we divided our initial triplet sample of 97 galaxy systems (see Section \ref{triplet_identification}) in three subsamples, corresponding to the percentiles of 33.33\% and 66.66\% of the redshift distribution. These redshift ranges correspond to 0.040$\le$z$<$0.060 ($S_1$), 0.060$\le$z$<$0.078 ($S_2$) and 0.078$\le$z$<$0.100 ($S_3$), respectively. In order to quantify any photometric redshift bias in the subsamples, we selected galaxies with spectroscopic redshift following three criteria: a) within 500 kpc of projected distance in the sky from the triplet centre, b) inside the velocity range of $|\Delta V|$=1000\ km\ s$^{-1}$ from the triplet centre, and c) brighter than $r$=17.77. It is important to mention that the subsamples include the triplet candidate members identified by the algorithm described in section \ref{triplet_identification}. This spectroscopic sample corresponds to 418 galaxies.

\begin{figure}
 \centering
 \includegraphics[scale=0.47]{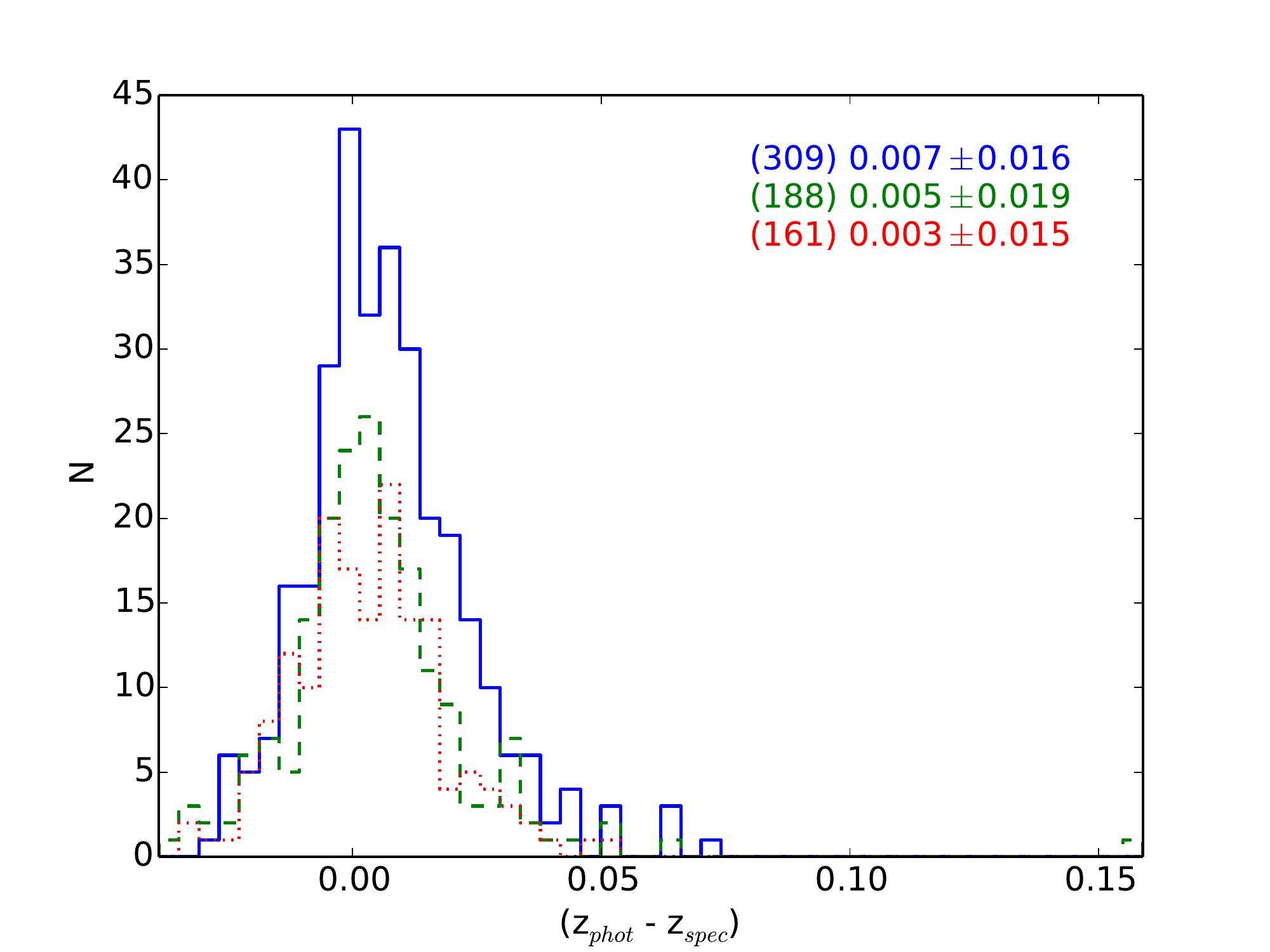}
 \caption{The redshift difference distributions between photometric and spectroscopic redshifts around triplet candidates for the redshift bins 0.040$\le$z$<$0.060 (solid line), 0.060$\le$z$<$0.078 (dashed line) and 0.078$\le$z$<$0.100 (dotted line). The mean and standard deviation of the redshift differences are at the top right of this figure. In parenthesis are the numbers of galaxies with spectroscopy around triplets in each redshift bin.}
 \label{fig_hist_deltaz}
\end{figure}

Figure \ref{fig_hist_deltaz} shows the distributions of the differences between the photometric and spectroscopic redshifts ($\Delta z=(z_{phot} - z_{spec})$) of galaxies with spectra around triplets. The mean difference of a given subsample, its mean photometric redshift bias, is taken into account to correct the photometric redshift of the i-th galaxy at the j-th redshift bin as: 

\begin{equation}
 z_{phot,corr,i} = z_{phot,i} - \mu_{\Delta z,j},
\end{equation}
where $z_{phot,i}$ is the photometric redshift of the i-th galaxy in the SDSS database and $\mu_{\Delta z,j}$ is the mean bias of the j-th redshift subsample. This redshift bias is around $10^{-3}$ in all redshift bins, a value comparable to the velocity interval used to define if a galaxy belongs to a triplet or not, showing that this correction is necessary to evaluate triplet membership.

Next, triplet member candidates are selected only adopting objects with photometric redshifts fullfilling criterion (a) above and within  $\pm\sigma_{\Delta z}$, the standard deviation of the bias between the photometric and spectroscopic redshifts from the triplet redshift bin. If the corrected photometric redshift of a galaxy candidate is contained in the redshift range of the triplet and its absolute magnitude at $r$-band is brighter than $-20.5 + 5 \log h_{70}$, it indicates that the candidate belongs to the triplet, i.e., the triplet candidate is, actually, a more complex system and is excluded from our triplet sample. In this candidate member selection, the dispersion between the spectroscopic and photometric redshifts is used instead of the mean SDSS photometric redshift uncertainty. Our choice represents more suitable values for the magnitude range adopted. The root mean square error (rmse) found by the SDSS group for the SDSS/DR8 spectroscopic sample is quite similar to the dispersions presented here, being 0.018\footnote{https://www.sdss3.org/dr10/algorithms/photo-z.php}.

Our initial sample of 97 triplet candidates is reduced of 14 objects that present one or more additional galaxies in the $\pm \sigma_{\Delta z}$ range. The numbers of systems excluded are 6, 3 and 5 in the redshift bins $S_1$, $S_2$ and $S_3$, respectively. The additional galaxies considered members of the systems have corrected absolute magnitudes (see section \ref{data}) in $r$-band between -20.54 and -22.65. In addition, we also excluded three triplets which harbour galaxies visually classified as Seyfert I (see section \ref{triplet_identification}). Thus, after the spectroscopic incompleteness analysis and the Seyfert I exclusion, our final triplet sample comprises 80 systems.

\label{lastpage}

\end{document}